\documentclass[sigconf]{acmart}

\usepackage{tabularx}
\newcolumntype{Y}{@{}>{\centering\arraybackslash}X@{}}
\usepackage{multirow}
\usepackage{enumitem}
\usepackage{subcaption}
\usepackage[ruled,linesnumbered]{algorithm2e}

\usepackage{amsmath,amsfonts,bm}

\def\Tabref#1{Table~\ref{#1}}

\def\Figref#1{Figure~\ref{#1}}

\def\Secref#1{Section~\ref{#1}}

\def\eqref#1{equation~\ref{#1}}
\def\Eqref#1{Equation~\ref{#1}}

\def\Algref#1{Algorithm~\ref{#1}}

\def\1{\bm{1}}

\def\eps{{\epsilon}}

\def\vc{{\bm{c}}}

\def\vs{{\bm{s}}}

\def\vu{{\bm{u}}}

\def\vx{{\bm{x}}}
\def\vy{{\bm{y}}}
\def\vz{{\bm{z}}}

\def\mA{{\bm{A}}}

\def\mD{{\bm{D}}}

\def\mF{{\bm{F}}}

\def\mI{{\bm{I}}}

\def\mO{{\bm{O}}}

\def\mU{{\bm{U}}}

\def\mW{{\bm{W}}}
\def\mX{{\bm{X}}}

\DeclareMathAlphabet{\mathsfit}{\encodingdefault}{\sfdefault}{m}{sl}
\SetMathAlphabet{\mathsfit}{bold}{\encodingdefault}{\sfdefault}{bx}{n}

\def\gI{{\mathcal{I}}}

\def\gL{{\mathcal{L}}}

\def\gN{{\mathcal{N}}}
\def\gO{{\mathcal{O}}}

\def\gU{{\mathcal{U}}}

\def\gX{{\mathcal{X}}}

\newcommand{\E}{\mathbb{E}}

\newcommand{\R}{\mathbb{R}}

\copyrightyear{2024}
\acmYear{2024}
\setcopyright{acmlicensed}\acmConference[SIGIR '24]{Proceedings of the 47th International ACM SIGIR Conference on Research and Development in Information Retrieval}{July 14--18, 2024}{Washington, DC, USA}
\acmBooktitle{Proceedings of the 47th International ACM SIGIR Conference on Research and Development in Information Retrieval (SIGIR '24), July 14--18, 2024, Washington, DC, USA}
\acmDOI{10.1145/3626772.3657759}
\acmISBN{979-8-4007-0431-4/24/07}

\begin{document}

\title{Graph Signal Diffusion Model for Collaborative Filtering}

\author{Yunqin Zhu}
\affiliation{
    \institution{University of Science and Technology of China}
    \department{School of Information Science and Technology}
    \city{Hefei}
    \country{China}}
\email{haaasined@gmail.com}
\orcid{0009-0002-9709-2417}

\author{Chao Wang}
\authornote{Corresponding authors: Chao Wang and Hui Xiong.}
\affiliation{
    \department{Guangzhou HKUST Fok Ying Tung Research Institute}
    \city{Guangzhou}
    \country{China}
}
\affiliation{
    \institution{University of Science and Technology of China}
    \department{School of Computer Science and Technology}
    \city{Hefei}
    \country{China}
}
\email{chadwang2012@gmail.com}
\orcid{0000-0001-7717-447X}

\author{Qi Zhang}
\affiliation{
    \institution{Shanghai AI Laboratory}
    \city{Shanghai}
    \country{China}}
\email{zhangqi.fqz@gmail.com}
\orcid{0000-0003-2942-7430}

\author{Hui Xiong}
\authornotemark[1]
\affiliation{
    \institution{The Hong Kong University of Science and Technology (Guangzhou), Thrust of Artificial Intelligence}
    \city{Guangzhou}
    \country{China}
}
\affiliation{
    \institution{The Hong Kong University of Science and Technology}
    \department{Department of Computer Science and Engineering}
    \city{Hong Kong SAR}
    \country{China}
}
\email{xionghui@ust.hk}
\orcid{0000-0001-6016-6465}

\begin{abstract}
    Collaborative filtering is a critical technique in recommender systems. It has been increasingly viewed as a conditional generative task for user feedback data, where newly developed diffusion model shows great potential. However, existing studies on diffusion model lack effective solutions for modeling implicit feedback. Particularly, the standard isotropic diffusion process overlooks correlation between items, misaligned with the graphical structure of the interaction space. Meanwhile, Gaussian noise destroys personalized information in a user's interaction vector, causing difficulty in its reconstruction. In this paper, we adapt standard diffusion model and propose a novel {\itshape \textbf{G}raph Signal D\textbf{iff}usion Model for \textbf{C}ollaborative \textbf{F}iltering} (named GiffCF). To better represent the correlated distribution of user-item interactions, we define a generalized diffusion process using heat equation on the item-item similarity graph. Our forward process smooths interaction signals with an advanced family of graph filters, introducing the graph adjacency as beneficial prior knowledge for recommendation. Our reverse process iteratively refines and sharpens latent signals in a noise-free manner, where the updates are conditioned on the user's history and computed from a carefully designed two-stage denoiser, leading to high-quality reconstruction. Finally, through extensive experiments, we show that GiffCF effectively leverages the advantages of both diffusion model and graph signal processing, and achieves state-of-the-art performance on three benchmark datasets.
\end{abstract}



\maketitle

\section{Introduction}

\begin{figure*}
    \centering
    \includegraphics[width=0.85\textwidth]{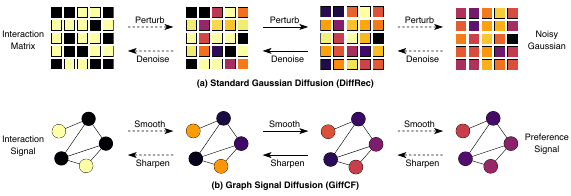}\label{fig:giffcf}
    \caption{Comparison of different diffusion processes for implicit feedback. Standard Gaussian diffusion destroys personalized information in the interaction matrix with Gaussian noise. On the contrary, our graph signal diffusion corrupts interaction signals into smoothed preference signals, leveraging the graphical structure of the interaction space. Each square in (a) represents a matrix entry, and each circle in (b) represents a scalar-valued graph node. Deeper colors indicate higher values.
    }\label{fig:conditional-diffusion}
    \Description{In subfigure (a) of standard Gaussian diffusion (DiffRec), the interaction matrix appears to be a grid of squares with two distinct colors. It gradually becomes a noisy Gaussian after being perturbed in multiple steps, which is represented by a grid of squares with mixed colors. The noisy Gaussian is then recovered back to the interaction matrix by denoising in multiple steps. In subfigure (b) of graph signal diffusion (GiffCF), the interaction signal appears to be an undirected graph with two distinct node colors. It gradually becomes a preference signal after being smoothed in multiple steps, which is represented by a graph of nodes with mixed colors. The preference signal is then recovered back to the interaction signal by sharpening in multiple steps.}
\end{figure*}

Collaborative filtering (CF) serves as a critical technique in recommender systems. It aims to reveal a user' hidden preferences through all the observed user feedback \cite{huCollaborativeFilteringImplicit2008}. Over the past few decades, researchers have designed various neural networks to mine collaborative patterns from implicit feedback, such as graph neural networks (GNNs) \cite{wangNeuralGraphCollaborative2019,heLightGCNSimplifyingPowering2020,fanGraphTrendFiltering2022a} and autoencoders (AEs) \cite{wuCollaborativeDenoisingAutoEncoders2016,liangVariationalAutoencodersCollaborative2018,maLearningDisentangledRepresentations2019,wangPersonalizedExplainableEmployee2021}. Some of these methods construct pair- \cite{rendleBPRBayesianPersonalized2009} or set-wise \cite{wangSetRankSetwiseBayesian2023} ranking loss to learn the partial order relations behind preference scores. Another increasingly popular paradigm is to minimize the point-wise distance between a reconstructed user interaction vector and its ground truth. Related works often perform dropout on the input interactions, thereby encouraging the model to predict missing ones \cite{wuCollaborativeDenoisingAutoEncoders2016}. This paradigm essentially regards CF as an inverse problem, which requires estimating the distribution of a user's potential interactions given its masked history. Many models have been proposed for this conditional generative task. For example, Mult-VAE \cite{liangVariationalAutoencodersCollaborative2018} assumed that interaction vectors follow a multinomial distribution and used variational inference to robustly optimize for likelihood. Another line of works \cite{wangGraphGANGraphRepresentation2018,wangIRGANMinimaxGame2017,chaeCFGANGenericCollaborative2018} incorporated adversarial training to produce more realistic interactions. More recently, DiffRec \cite{wangDiffusionRecommenderModel2023} treated interaction vectors as noisy latent states in DDPM \cite{hoDenoisingDiffusionProbabilistic2020}, pioneering the application of diffusion model to recommendation.

Diffusion model (DM) have made remarkable successes in solving inverse problems across multiple domains, such as image inpainting \cite{rombachHighResolutionImageSynthesis2022}, speech enhancement \cite{luConditionalDiffusionProbabilistic2022}, and time series imputation \cite{tashiroCSDIConditionalScorebased2021}. The ability of DM to learn complex distributions can be attributed to its hierarchical structure, which allows iterative updates of the latent states toward fine-grained directions. However, for the task of CF, the performance of standard DM is still far from satisfactory, especially when compared with the state-of-the-art graph signal processing techniques \cite{shenHowPowerfulGraph2021,fuRevisitingNeighborhoodbasedLink2022,choiBlurringSharpeningProcessModels2023} on large datasets (see \Secref{sec:comparison}). We argue that standard Gaussian diffusion could have limited capabilities when modeling implicit feedback, primarily due to two reasons: (1) the isotropic diffusion process overlooks the correlation between items, leading to a misalignment with the graphical structure of the interaction space; (2) Gaussian noise destroys personalized information in user interaction vectors, causing difficulty in their reverse reconstruction. For the latter issue, \citet{wangDiffusionRecommenderModel2023} proposed to limit the variance of added noise. However, their approach still suffers from the first issue, leaving the design space of DM largely unexplored.

In this work, we adapt standard DM to address the above issues, and propose a novel {\itshape \textbf{G}raph Signal D\textbf{iff}usion Model for \textbf{C}ollaborative \textbf{F}iltering} (named GiffCF). Seeing that user-item interactions are high-dimensional, sparse, yet correlated in nature, we break the routine of adding or removing Gaussian noise for distribution modeling, as image DMs usually do \cite{hoDenoisingDiffusionProbabilistic2020}. Instead, we view interactions as signals on the item-item similarity graph, and then define a generalized diffusion process that simulates graph heat equation (see \Figref{fig:conditional-diffusion}). Inspired by graph signal processing baselines, we design a novel family of graph smoothing filters to corrupt interaction signals in the forward process. Such forward filters introduce the graph adjacency as helpful prior knowledge for the recommendation task, overcoming the first challenge. As for high-quality reconstruction, our reverse process iteratively refines and sharpens preference signals in a noise-free manner, where the update direction is conditioned on the user history and computed from a carefully designed two-stage denoiser. Through extensive experiments, we show that GiffCF effectively leverages the advantages of both DM and graph signal processing, setting new records on three benchmark datasets. Our code is available at \url{https://github.com/HasiNed/GiffCF}. To summarize, our contributions are:

\begin{enumerate}[leftmargin=*]
    \item We propose a generalized diffusion process for CF that leverages the structure of the interaction space by smoothing and sharpening signals on the item-item similarity graph.
    \item We unify graph signal processing baselines and propose a novel family of graph filters to smooth interaction signals in the forward process and introduce helpful inductive bias for CF.
    \item We conduct comprehensive experiments on multiple benchmark datasets to show the superiority of our method over competitive baselines. We also provide empirical analysis of both the forward and reverse processes.
\end{enumerate}


\section{Background}\label{sec:background}

\paragraph{Conditional Gaussian diffusion}

DM \cite{sohl-dicksteinDeepUnsupervisedLearning2015,hoDenoisingDiffusionProbabilistic2020,kingmaVariationalDiffusionModels2021} has achieved state-of-the-art in many conditional generative tasks. Given an initial sample $\vx$ from the data distribution $q(\vx)$, the forward process of standard Gaussian diffusion is defined through an increasingly noisy sequence of latent variables deviating from $\vx$, written as
\begin{equation}\label{eq:diff}
    \begin{gathered}
        \vz_t = \alpha_t \vx + \sigma_t \bm{\eps}_t, \\
        \bm{\eps}_t \sim \gN(\bm{0}, \bm{I}),\quad t=0,1,2,\dots,T,
    \end{gathered}
\end{equation}
or $q(\vz_t|\vx) = \gN(\vz_t; \alpha_t \vz, \sigma_t^2\bm{I})$. In a variance-preserving noise schedule, the noise level $\sigma_t$ monotonically increases from 0 to 1, while the signal level $\alpha_t$ is constrained by $\alpha_t^2 + \sigma_t^2 = 1$. To see how $\vz_s$ and $\vz_t$ from two arbitrary timesteps transition to each other, we write $q(\vz_s|\vz_t, \vx) = \gN(\vz_s; \alpha_s \vx + \sqrt{\sigma_s^2 - \sigma_{s|t}^2} \bm{\eps}_t, \sigma_{s|t}^2\bm{I})$. Here, $\sigma_{s|t}^2$ depends on further assumptions. For instance, a Markov forward process leads to $\sigma_{t|s}^2 = \sigma_t^2 - \frac{\alpha_t^2\sigma_s^2}{\alpha_s^2}$ and $\sigma_{s|t}^2 = \sigma_s^2 - \frac{\alpha_t^2\sigma_s^4}{\alpha_s^2\sigma_t^2},\, \forall s < t$. The reverse process is a parameterized hierarchical model, from which we can sample $\vx$ given some condition $\vc$:
\begin{equation}\label{eq:reverse}
    p_\theta(\vx|\vc) = \int p_\theta(\vx|\vz_0)\left[\prod_{t=1}^{T}p_\theta(\vz_{t-1}|\vz_t,\vc)\right] p_\theta(\vz_T|\vc) \mathrm{d} \vz_{0:T},
\end{equation}
In a simplified setup, $p_\theta(\vx|\vz_0)$ is usually chosen to be an identity mapping $p(\vx|\vz_0) = \delta(\vx - \vz_0)$, and $p_\theta(\vz_T|\vc)$ to be a unit Gaussian distribution $p(\vz_T) = \gN(\vz_T; \bm{0}, \bm{I})$. The remaining terms are approximated using the transition relations we derived above, \emph{i.e.}, $p_{\theta}(\vz_{t-1}|\vz_t,\vc) = q(\vz_{t-1}|\vz_t, \vx = \hat{\vx}_\theta(\vz_t, \vc, t))$, $t=1,\dots,T$. Here, $\hat{\vx}_\theta(\vz_t, \vc, t)$ is a learnable denoising network, or {\itshape denoiser}, that is trained to maximize the lower bound of log-likelihood. Typically, the diffusion loss is reduced to the following form:
\begin{equation}\label{eq:diffusion-loss}
    \gL = \E_{t\sim \gU(1,T), q(\vz_t| \vx)}[w_t\|\hat{\vx}_\theta(\vz_t, \vc, t) - \vx\|^2],
\end{equation}
where $w_t$ is a weighting factor that balances the gradient magnitude of different timesteps. Intuitively, the reverse process is composed of a series of denoising steps, each of which updates the latent state $\vz_t$ toward a direction with higher conditional likelihood, and finally arrives at a high-quality sample of $\vx$.

\paragraph{CF as an inverse problem}

\begin{figure}
    \centering
    \begin{subfigure}{\columnwidth}
        \centering
        \includegraphics[width=0.6\textwidth]{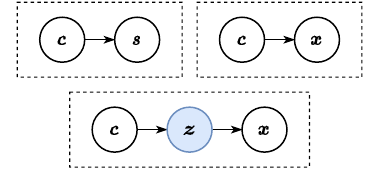}
        \caption{Non-diffusion recommender models. Instead of directly predicting scores $\vs$ (top-left), DAE (top-right) and VAE (bottom) learn to reconstruct interactions $\vx$.
        }\label{fig:non-diff}
    \end{subfigure}
    \begin{subfigure}{\columnwidth}
        \centering
        \includegraphics[width=0.8\textwidth]{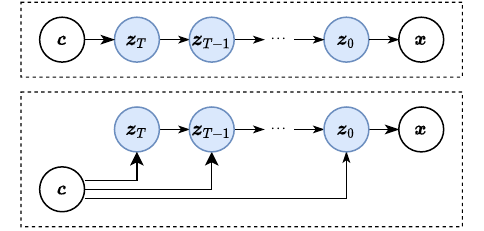}
        \caption{Diffusion recommender models. Recent works studied unconditional DMs for CF (top). We incorporate a conditioning mechanism of historical interactions $\vc$ (bottom).}\label{fig:diff}
    \end{subfigure}
    \caption{Graphical models for different recommender models at inference time. Latent variables are colored in blue.}
    \Description{The subfigure (a) contains 3 non-diffusion recommender models: in the top left one, variable c directly points to a variable s; in the top right one, or DAE, a variable c points to a variable x; in the bottom one, or VAE, a variable c points to a latent variable z colored in blue, which then points to a variable x. 
    The subfigure (b) contains 2 diffusion recommender models: in the top one, or previous unconditional DMs for CF, a variable c points to a chain of latent variables z_t colored in blue, with the last one pointing to x; in the bottom one, or our proposed method with a conditioning mechanism, there is a chain of latent variables z_t colored in blue that finally points to x, with an additional variable c pointing to each latent variable.}
\end{figure}

Similar to image inpainting \cite{rombachHighResolutionImageSynthesis2022} or time series imputation \cite{tashiroCSDIConditionalScorebased2021}, CF can be formulated as an inverse problem and solved using a conditional generative model such as VAE or DM. Denote the set of users as $\gU$ and the set of items as $\gI$. For each user in $\gU$, we aim to rank all the items in $\gI$ by predicting a score vector $\hat{\vs}\in\R^{|\gI|}$, and then recommend top-$K$ plausible items to fulfill the user's needs. In the setting of CF with implicit feedback, our supervision is a user-item interaction matrix $\mX\in \{0,1\}^{|\gU|\times|\gI|}$ with each row $\vx_u\in\{0,1\}^{|\gI|}$ denoting the interaction vector of the user $u$. The user $u$ has interacted with item $i$ if $x_{u,i} = 1$, whereas no such interaction has been observed if $x_{u,i} = 0$. The difficulty we encounter here is that the true score vector $\vs$ is not directly available from the binary entries of $\mX$. To address this issue, we can follow the paradigm of masked autoencoders (or denoising autoencoders, DAE) as in \cite{wuCollaborativeDenoisingAutoEncoders2016}: (1) for each training interaction vector $\vx$, we randomly drop out some of its components to obtain a degraded version $\vc$, and then train an inverse model $p_\theta(\vx|\vc)$ to reconstruct $\vx$ from $\vc$; (2) at inference time, we obtain the predicted scores $\hat{\vs}$ by reconstructing from $u$'s historical interactions $\vx_u$. In this way, we implicitly model the expectation $\E[\vx|\vc=\vx_u]$ as the ground-truth scores $\vs$. Subsequent works have continued this paradigm, such as Mult-VAE \cite{liangVariationalAutoencodersCollaborative2018}, which learned a multinomial $p(\vx|\vc)$ by means of variational Bayes. In this paper, we are interested in learning a hierarchical $p(\vx|\vc)$ using conditional DM.

\paragraph{Unconditional DM for CF}

Before diving into our method, we point out that unconditional DM can be used to solve a certain kind of inverse problems. Specifically, we can model the condition itself as some latent state $\vz_t$ corrupted by the forward process. Taking CF as an example, \citet{wangDiffusionRecommenderModel2023} assumed an intrinsic Gaussian noise in each $\vx_u$, and reconstructed interactions from $p_\theta(\vx|\vz_T=\vx_u)$. Their method instantiates \Eqref{eq:reverse} with $p(\vz_T|\vc) = \delta(\vz_T - \vc)$ and $p_\theta(\vz_{t-1}|\vz_t,\vc) = q(\vz_{t-1}|\vz_t, \vx = \hat{\vx}_\theta(\vz_t, t))$, $t=1,\dots,T$. Notably, they still employed dropout for all the latents $\vz_t$ during training. As a result, their denoiser $\hat{\vx}_\theta(\vz_t, t)$ is forced to invert both the dropout noise and a small-scale Gaussian noise, which can be understood as a robustness-enhanced version of masked autoencoder. In this paper, we decouple the two types of noise and compute each reverse step with user history $\vc$ as an additional condition. The difference between our modeling method and theirs is illustrated in \Figref{fig:diff}.

\section{Proposed Method}

In this section, we elucidate our modifications of both the forward and reverse processes, leading to a novel DM for CF named GiffCF.

\subsection{Graph Signal Forward Process}

The high-dimensional and sparse nature of implicit feedback makes it difficult to model the whole distribution of interactions using Gaussian diffusion. Fortunately, recent works have shown that the forward corruption manner of DM is not unique \cite{darasSoftDiffusionScore2022,bansalColdDiffusionInverting2023,rissanenGenerativeModellingInverse2022,hoogeboomBlurringDiffusionModels2022}. In the context of recommender systems, our target is to accurately fulfill a user's needs with predicted items, which is only relevant to $\E[\vx|\vc=\vx_u]$. Therefore, we hypothesize that it is neither efficient nor necessary to rely on Gaussian perturbations to fully explore the distribution of interaction vectors. Rather, we are interested in finding a forward process tailored for implicit feedback data, introducing inductive bias beneficial for CF.

\subsubsection{Graph Smoothing with Heat Equation}

A user’s interaction vector can be considered a signal on the item-item similarity graph. Inspired by the over-smoothing artifact in GNNs \cite{ruschSurveyOversmoothingGraph2023}, we immediately notice a corruption manner dedicated to graph signals: repeating smoothing operations until it reaches a steady state. Interestingly, similar attempts using blurring filters for forward diffusion have been made in image synthesis literature \cite{rissanenGenerativeModellingInverse2022,hoogeboomBlurringDiffusionModels2022}. The general idea of a smoothing process is to progressively exchange information on a (discretized) Riemanian manifold (\emph{e.g.}, an item-item graph), described by the following continuous-time PDE:
\begin{equation}\label{eq:heat}
    \frac{\partial \vz} {\partial \tau} = \alpha \nabla^2 \vz,
\end{equation}
known as the \emph{heat equation}. Intuitively, the Laplacian operator $\nabla^2$ measures the information difference between a point and the average of its neighborhood, and $\alpha$ captures the rate at which information is exchanged over time $\tau$. For interaction signals, we let $\nabla^2 = \mA - \mI$ by convention \cite{shenHowPowerfulGraph2021}, where $\mA$ is a normalized adjacency matrix of the item-item graph. It is straightforward to verify the closed form solution of \Eqref{eq:heat} with some initial value $\vz(0)$:
\begin{equation}\label{eq:solution}
    \begin{aligned}
        \vz(\tau) & = \exp\{-\tau\alpha (\mI - \mA)\} \vz(0)                                                 \\
                  & = \mU \exp\{-\tau\alpha (\bm{1} - \bm{\lambda})\} \odot \mU^\top \vz(0),\quad \tau\ge 0,
    \end{aligned}
\end{equation}
where $\mU = [\vu_1, \dots, \vu_{|\gI|}]$ is the matrix of unit eigenvectors of $\mA$ with the corresponding eigenvalues $\bm{\lambda} = [\lambda_1, \dots, \lambda_{|\gI|}]$ sorted in descending order, and $\odot$ denotes Hadamard product. The orthonormal matrix $\mU^\top$ is often referred to as \emph{graph Fourier transform} (GFT) matrix and $\mU$ as its inverse, and the eigenvalues of $-\nabla^2$, \emph{i.e.} $\bm{1} - \bm{\lambda}$, are called \emph{graph frequencies} \cite{chungSpectralGraphTheory}. Thus, \Eqref{eq:solution} can be interpreted as exponentially decaying each high-frequency component of the latent signal $\vz$ with an anisotropic rate $\alpha(\bm{1} - \bm{\lambda})$. As $\tau \to +\infty$, the signal $\vz(\tau)$ converges to an over-smoothed steady state $\vz(+\infty) = \vu_1\vu_1^\top\vz(0)$.

A problem of this naive smoothing process is the computational intractability of finding the matrix exponential or fully diagonalizing $\mA$ for item-item graphs with thousands and millions of nodes. One solution is to approximate the continuous-time filter in \Eqref{eq:solution} using its first-order Taylor expansion with respect to $\tau$:
\begin{equation}
    \exp\{-\tau\alpha (\mI - \mA)\} = (1-\tau\alpha )\mI + \tau\alpha \mA + \gO(\tau).
\end{equation}
Here, we simply set $\tau = 1$ for our final state and linearly interpolate the intermediate states to avoid computational overhead, yielding a sequence of forward filters
\begin{equation}\label{eq:filter}
    \mF_t = (1-\tau_t\alpha )\mI + \tau_t\alpha \mA,\quad t=0,1,2,\dots,T,
\end{equation}
where $0 = \tau_0 < \tau_1 < \dots < \tau_T = 1$ defines a \emph{smoothing schedule}. By multiplying $\mF_t$ with the initial interaction signal $\vx$, we obtain a sequence of smoothed preference signals $\vz_t$ as the diffusion latent states in the forward process.

Note that by first-order truncation, we only consider 1-hop propagation of message on the item-item graph and do not actually reach the steady state. However, this is sufficient for our purpose, since an over-smoothed latent state may lose personalized information and make marginal improvement to recommendation performance. Unlike the standard Gaussian diffusion that adds noise and destroys useful signals, our forward process itself has the capability of reducing dropout noise and introduces the structure of the interaction space as helpful prior knowledge. This coincides with the idea of graph signal processing techniques for CF \cite{shenHowPowerfulGraph2021,fuRevisitingNeighborhoodbasedLink2022,choiBlurringSharpeningProcessModels2023}. We will later draw inspiration from two strong baselines in this field and design the adjacency matrix $\mA$ in \Secref{sec:adjacency}.

\subsubsection{Anisotropic Diffusion in the Graph Spectral Domain}

So far, we have introduced a deterministic corruption manner for interaction signals but ignored the probabilistic modeling aspect of DM. To better understand our forward process, we add a Gaussian noise term similar to standard diffusion, yielding the complete formulation of \emph{graph signal forward process}:
\begin{equation}\label{eq:graph-forward}
    \begin{gathered}
        \vz_t = \mF_t\vx + \sigma_t\bm{\eps}_t, \\ \bm{\eps}_t \sim \gN(\bm{0}, \bm{I}),\quad t=0,1,2,\dots,T,
    \end{gathered}
\end{equation}
where $(\mF_t)_{t=1}^T$ is defined in \Eqref{eq:filter} and $(\sigma_t)_{t=1}^T$ controls the noise level. Comparing it with \Eqref{eq:diff}, we can see that the mere difference between standard diffusion and ours lies in the choice of $\mF_t$: the former uses a diagonal matrix $\mF_t = \alpha_t\mI$ that isotropically decays the interactions, while we adopt a smoothing filter to exchange information on the item-item graph. 

To give a further correspondence between the two processes, we diagonalize our filters as
\begin{equation}\label{eq:filter-spectral}
    \mF_t = \mU[(1-\tau_t\alpha )\bm{1} + \tau_t\alpha \bm{\lambda}]\odot\mU^\top,\quad t=0,1,2,\dots,T,
\end{equation}
Let the GFT of $\vx$, $\vz_t$, and $\bm{\eps}_t$ be $\tilde{\vx} = \mU^\top\vx$, $\tilde{\vz}_t = \mU^\top\vz_t$, and $\tilde{\bm{\eps}}_t = \mU^\top\bm{\eps}_t$, respectively. By plugging \Eqref{eq:filter-spectral} into \Eqref{eq:graph-forward}, we get an equivalent formulation of graph signal forward process in the graph spectral domain:
\begin{equation}
    \begin{gathered}
        \tilde{\vz}_t = [(1-\tau_t\alpha )\bm{1} + \tau_t\alpha \bm{\lambda}]\odot\tilde{\vx} + \sigma_t\tilde{\bm{\eps}}_t, \\
        \tilde{\bm{\eps}}_t \sim \gN(\bm{0}, \bm{I}),\quad t=0,1,2,\dots,T.
    \end{gathered}
\end{equation}
It turns out to be a special case of Gaussian diffusion with an anisotropic noise schedule; that is, the scalar signal level $\alpha_t$ in \Eqref{eq:diff} now becomes a vector signal level $\bm{\alpha}_t = (1-\tau_t\alpha )\bm{1} + \tau_t\alpha \bm{\lambda}$.

\subsubsection{Identifying the Item-Item Graph}\label{sec:adjacency}

The interaction matrix $\mX$ provides the adjacency between a user node and an item node on the user-item bipartite graph, yet to smooth an interaction vector, it is necessary to measure the similarity between two item nodes. Let us define $\mD_\gU = \mathrm{diagMat}(\mX\bm{1})$ and $\mD_\gI = \mathrm{diagMat}(\mX^\top\bm{1})$ as the degree matrices of users and items, respectively. A popular approach to construct an adjacency matrix for the item-item graph is to stack two convolution kernels $\tilde{\mA}_{\rm LGN}$ of LightGCN \cite{heLightGCNSimplifyingPowering2020}:
\begin{equation}
    \begin{aligned}
        \tilde{\mA}_{\rm LGN}^2 & =
        \begin{bmatrix}
            \mO                                                  & \mD_\gU^{-\frac{1}{2}}\mX\mD_\gI^{-\frac{1}{2}} \\
            \mD_\gI^{-\frac{1}{2}}\mX^\top\mD_\gU^{-\frac{1}{2}} & \mO
        \end{bmatrix}^2                                                                 \\
                                & = \begin{bmatrix}
                                        \mD_\gU^{-\frac{1}{2}}\mX\mD_\gI^{-1}\mX^\top\mD_\gU^{-\frac{1}{2}} & \mO                                                                  \\
                                        \mO                                                                 & \mD_\gI^{-\frac{1}{2}}\mX^\top\mD_\gU^{-1}\mX\mD_\gI^{-\frac{1}{2}}
                                    \end{bmatrix},
    \end{aligned}
\end{equation}
and then use the lower-right block, which we denote by $\mA_{\frac{1}{2}, 1, \frac{1}{2}}$. More recently, \citet{fuRevisitingNeighborhoodbasedLink2022} proposed a generalized form of two-step link propagation on the bipartite graph. Here, we reformulate it as an item-item similarity matrix
\begin{equation}\label{eq:linkprop}
    \mA_{\beta, \gamma, \delta} = \mD_\gI^{-\delta}\mX^\top\mD_\gU^{-\gamma}\mX\mD_\gI^{-\beta},
\end{equation}
where $\beta, \gamma, \delta$ are tunable parameters. For example, when $\beta=\gamma=\delta=0$, the similarity measure degrades into the number of common neighbors between two item nodes \cite{newmanClusteringPreferentialAttachment2001}. The logic of normalization lies in that nodes with higher degrees often contain less information about their neighbors, and thus should be down-weighted. In our preliminary experiments, we find that using an equal order of normalization on three user/item nodes, {\itshape i.e.} $\beta=\gamma=\delta=\frac{1}{2}$, works consistently better than the LightGCN-style stacking. Therefore, it serves as our default choice.

Nevertheless, this filter only aggregates information from 2-hop neighbors on the original bipartite graph. Breaking this limitation, we strengthen its expressiveness with an ideal low-pass filter similar to GF-CF \cite{shenHowPowerfulGraph2021}. Basically, if we denote $\mU_{\beta, \gamma, \delta, d}$ as the matrix of eigenvectors corresponding to the top-$d$ eigenvalues of $\mA_{\beta, \gamma, \delta}$ (lowest $d$ graph frequencies), the low-pass filter can be decomposed as $\mU_{\beta, \gamma, \delta, d}\mU_{\beta, \gamma, \delta, d}^\top$. Multiplying it with an interaction vector $\vx$ filters out high-frequency components while preserving high-order signals in the low-frequency subspace.

We summarize representative CF techniques related to graph filters in \Tabref{tab:filter}. Unifying the designs of LinkProp and GF-CF, we propose the following adjacency matrix in our forward filters:
\begin{equation}
    \mA = \frac{1}{1+\omega}\left(\mA_{\beta,\gamma,\delta}/\|\mA_{\beta,\gamma,\delta}\|_2 + \omega\mU_{\beta,\gamma,\delta,d}\mU_{\beta,\gamma,\delta,d}^\top\right),
\end{equation}
where $d$ and $\omega$ are the cut-off dimension and the strength of ideal low-pass filtering, respectively. Note that the matrix has been scaled to a unit norm.

\begin{table}
    \centering\small
    \setlength\extrarowheight{1pt}
    \caption{A summary of graph filters for interaction signals. $\vx$ denotes input and $\vy$ denotes filtered output. Scaling factors that do not affect top-$K$ recommendation results are omitted.}\label{tab:filter}
    \begin{minipage}{\columnwidth}
        {
            \begin{tabularx}{\textwidth}{l|l}
                \hline
                Method                                                              & Formulation                                                                                                                                             \\
                \hline
                Linear AE                                                           & $\vy = \mW_{\rm dec}\mW_{\rm enc}\vx$, $\mW_{\rm enc},\mW_{\rm dec}^\top \in \R^{|\gI|\times d}$                                                        \\
                \hline
                LinkProp \cite{fuRevisitingNeighborhoodbasedLink2022}               & $\vy = \mA_{\beta, \gamma, \delta}\vx = \mD_\gI^{-\delta}\mX^\top\mD_\gU^{-\gamma}\mX\mD_\gI^{-\beta} \vx$                                              \\
                \hline
                                                                                    & $\vy = \mA_{\rm HE}\vx = \mA_{\frac{1}{2}, 1, \frac{1}{2}} \vx$                                                                                         \\
                {GF-CF  \cite{shenHowPowerfulGraph2021}}                            & $\vy = \mA_{\rm IDL}\vx = \mD_\gI^{\frac{1}{2}}\mU_{\frac{1}{2}, 1, \frac{1}{2}, d}\mU_{\frac{1}{2}, 1, \frac{1}{2}, d}^\top\mD_\gI^{-\frac{1}{2}} \vx$ \\
                                                                                    & $\vy = (\mA_{\rm HE} + \omega\mA_{\rm IDL})\vx$, $\omega \ge 0$                                                                                         \\
                \hline
                BSPM$^\dag$ \cite{choiBlurringSharpeningProcessModels2023}          & $\vy = (\mI-\omega'\mA_{\rm HE})(\mA_{\rm HE} + \omega\mA_{\rm IDL})\vx$, $\omega' \ge 0$                                                               \\
                \hline
                GiffCF (Ours)                                                       & $\vy = (\mA_{\beta,\gamma,\delta}/\|\mA_{\beta,\gamma,\delta}\|_2 + \omega\mU_{\beta,\gamma,\delta,d}\mU_{\beta,\gamma,\delta,d}^\top)\vx$              \\
                \hline
            \end{tabularx}\\
        }
        \footnotesize $^\dag$Euler method, single blurring/sharpening steps, and early merge.
    \end{minipage}

\end{table}

\paragraph{A note on time complexity.}

The truncated eigendecomposition can be implemented in a preprocessing stage using efficient algorithms \cite{baglamaAugmentedImplicitlyRestarted2005,muscoRandomizedBlockKrylov2015}. When applying smoothing filters, the link propagation term requires sparse matrix multiplication and costs $\gO(|\gX|)$ time for each $\vx$, where $|\gX|$ is the number of observed interactions in the training data; the ideal low-pass term requires low-rank matrix multiplication and costs $\gO(|\gI|d)$ time. Both of them sidestep the need for $\gO(|\gI|^2)$ dense multiplication.

\subsection{Graph Signal Reverse Process}

As the counterpart of our forward process, we expect a {\itshape graph signal reverse process} that iteratively removes the effect of smoothing on the latent preference signals, and finally reconstructs the original interaction vector for recommendation. This is similar to standard diffusion which uses multiple denoising steps to remove Gaussian noise from the latent state. Since we only change the corruption manner of interaction signals, the reverse process of GiffCF is still defined by the hierarchical generative model in \Eqref{eq:reverse}. And the transition relation between two timesteps $t$ and $s$ can be written as
\begin{equation}\label{eq:transition}
    q(\vz_s|\vz_t, \vx) = \gN(\vz_s; \mF_s\vx + \sqrt{\sigma_s^2 - \sigma_{s|t}^2} \bm{\eps}_t, \sigma_{s|t}^2\bm{I}).
\end{equation}
In this section, we lay the theoretical foundation of this reverse process and detail the implementation of our inference procedure.

\subsubsection{Deterministic Sampler}

As we have discussed before, noise in the diffusion processes may destroy useful signals and undermine the recommendation performance. Also, the ultimate goal of GiffCF is to give a high-quality estimation of $\E[\vx|\vc]$, instead of sampling diversified data from the distribution $p_\theta(\vx|\vc)$ like in image synthesis. Due to these reasons, we choose to use a deterministic sampler for the reverse process similar to DDIM \cite{songDenoisingDiffusionImplicit2020}.

Recalling \Eqref{eq:reverse} and \Eqref{eq:transition}, we assume $\sigma_{t-1|t}^2=0$ for all $p_\theta(\vz_{t-1}|\vz_t, \vc) = q(\vz_{t-1}|\vz_t, \vx = \hat{\vx}_\theta(\vz_t, \vc, t))$, $t=1,\dots,T$, making each transition from $\vz_t$ to $\vz_{t-1}$ deterministic:
\begin{equation}
    \vz_{t-1} = \mF_{t-1}\hat{\vx}_\theta + \frac{\sigma_{t-1}}{\sigma_t}(\vz_t - \mF_t\hat{\vx}_\theta),\quad t=1,\dots,T,
\end{equation}
where $\hat{\vx}_\theta(\vz_t, \vc, t)$ is the denoiser output at the specific timestep $t$.

For the prior distribution $p(\vz_T|\vc)$, the optimal decision should place all the probability mass on the ground-truth $\vz_T$ defined by the forward filter. However, this is not feasible in practice, since we have no access to the true interaction vector $\vx$ during inference. To compromise, we approximate $\vz_T$ by the smoothed version of historical interactions, \emph{i.e.} $\mF_T\vc$, because $\vx$ and $\vc$ should be close enough in the low-frequency subspace, which is exactly the foundational assumption of graph signal processing techniques for CF \cite{shenHowPowerfulGraph2021}. The resulting reverse process then becomes a completely deterministic mapping from $\mF_T\vc$ to $\vx$ (see \Algref{alg:inference}).

\subsubsection{Refining-Sharpening Decomposition}\label{sec:refine-sharpen}

To deepen our understanding of the reverse process, we decompose the update rule of $\vz_t$ into a refining term and a sharpening term:
\begin{equation}\label{eq:refine-sharpen-decomposition}
    \begin{aligned}
        \vz_{t-1} & = \left[\mF_{t} - (\tau_t - \tau_{t-1})\alpha(\mA-\mI)\right]\hat{\vx}_\theta + \frac{\sigma_{t-1}}{\sigma_t}(\vz_t - \mF_t\hat{\vx}_\theta)                                                                             \\
                  & = \vz_t + \underbrace{\left(1-\frac{\sigma_{t-1}}{\sigma_t}\right)(\mF_t\hat{\vx}_\theta - \vz_t)}_{\text{Refining Term}} + \underbrace{(\tau_t - \tau_{t-1})\alpha (\mI-\mA)\hat{\vx}_\theta}_{\text{Sharpening Term}}.
    \end{aligned}
\end{equation}
Intuitively, the refining term corrects the latent signal $\vz_t$ from the previous timestep by a weighted average of $\mF_t\hat{\vx}_\theta$ and $\vz_t$. By tuning the {\itshape noise decay} $\frac{\sigma_{t-1}}{\sigma_t}$, we can adjust the amount of refinement at each reverse step. This term can probably ease the negative effect of prior misspecification caused by setting $\vz_T = \mF_T\vc$. Meanwhile, the sharpening term highlights the difference of $\hat{\vx}_0$ from its smoothed version $\mA\hat{\vx}_0$, equivalently solving the heat equation in the reverse direction. In our implementation, we use a linear smoothing schedule with $\tau_t = t/T$, $t=0, 1,\dots,T$, and a constant noise decay $\frac{\sigma_{t-1}}{\sigma_t}$, $t=1,\dots,T$ as a hyper-parameter to balance the two terms.

\subsubsection{Derivation of the Optimization Objective}

The denoiser $\hat{\vx}_\theta$ plays an important role in the reverse process as all the update directions are computed from its output. In order to learn an effective denoiser for high-quality reconstruction, we try to maximize the conditional log-likelihood of the generated data:

\begin{equation}
    \begin{aligned}
         & \log p_\theta(\vx|\vc) = \log \E_{q(\vz_{0:T}|\vx)}\left[\frac{p_\theta(\vz_{0:T},\vx|\vc)}{q(\vz_{0:T}| \vx)}\right]                                                                               \\
         & \ge \E_{q(\vz_{0:T}|\vx)}\left[\log \frac{p_\theta(\vz_{0:T},\vx|\vc)}{q(\vz_{0:T}| \vx)}\right]                                                                                                    \\
         & = -\underbrace{D_{\mathrm{KL}} \left(q(\vz_T| \vx) \parallel p(\vz_T|\vc)\right)}_{\text{Prior Loss}} - \underbrace{\E_{q(\vz_0|\vx)}\left[-\log p(\vx| \vz_0)\right]}_{\text{Reconstruction Loss}} \\
         & - \underbrace{\sum^{T}_{t=1}\E_{q(\vz_t|\vx)} \left[D_{\mathrm{KL}} \left(q(\vz_{t-1}| \vz_t,\vx) \parallel p_\theta(\vz_{t-1}| \vz_t,\vc)\right)\right]}_{\text{Diffusion Loss}}.
    \end{aligned}
\end{equation}

The right-hand side of the above inequality is known as the \emph{variational lower bound} (VLB) \cite{kingmaVariationalDiffusionModels2021} of log-likelihood. Since we set $p(\vz_T|\vc)=\delta(\vz_T-\mF_T\vc)$ and $p(\vx|\vz_0)=\delta(\vx-\vz_0)$ in practice, the prior loss and reconstruction loss are both constant and can be ignored during optimization. We will now derive an estimator for the diffusion loss. Since $q(\vz_{t-1}| \vz_t,\vx)$ and $p_\theta(\vz_{t-1}| \vz_t,\vc)$ are both Gaussian, the KL divergence can be computed in a closed form:
\begin{equation}\label{eq:denoising}
    \begin{aligned}
         & \quad D_{\mathrm{KL}} \left(q(\vz_s| \vz_t,\vx) \parallel p_\theta(\vz_s| \vz_t,\vc)\right)                                                                                       \\
         & = \frac{1}{2\sigma^2_{s|t}}\|\mF_s(\hat{\vx}_\theta - \vx) + \sqrt{\sigma_s^2 - \sigma_{s|t}^2}(\hat{\bm{\eps}}_\theta - \bm{\eps}_t)\|^2.                                        \\
         & = \frac{1}{2\sigma^2_{s|t}}\left\|\left(\mF_s - \sqrt{\sigma_s^2 - \sigma_{s|t}^2}/\sigma_t \cdot\mF_t\right)(\hat{\vx}_\theta - \vx)\right\|^2.                                  \\
         & \le \frac{1}{2\sigma^2_{s|t}}\left\|\mF_s - \sqrt{\sigma_s^2 - \sigma_{s|t}^2}/\sigma_t \cdot\mF_t\right\|^2\|\hat{\vx}_\theta - \vx\|^2  = C_{s|t} \|\hat{\vx}_\theta - \vx\|^2,
    \end{aligned}
\end{equation}
where $\hat{\bm{\eps}}_\theta = (\vz_t - \mF_t\hat{\vx}_\theta)/\sigma_t$ . The inequality is due to the compatibility between the matrix and vector norms. Therefore, the total loss reduces to a summation of $T$ weighted square errors. Following the standard practice of DM, we sample the timestep $t$ uniformly at random and minimize a reweighted version of the diffusion loss, which is identical to the loss of standard DM (see \Eqref{eq:diffusion-loss}).

While image DMs typically learn a U-Net \cite{ronnebergerUNetConvolutionalNetworks2015} denoiser for predicting $\bm{\eps}_t$ and essentially set $w_t$ to the signal-to-noise ratio $\frac{\alpha_t^2}{\sigma_t^2}$ \cite{kingmaVariationalDiffusionModels2021}, in this paper, we choose to directly predict $\vx$ with $w_t=1$ for all $t=1,\dots, T$, as we find it more stable when used in conjunction with our denoiser architecture.

{
    \IncMargin{1.5em}
    \begin{algorithm}[t]
        \caption{GiffCF training.}\label{alg:training}
        \Indm
        \KwIn{Interaction matrix $\mX$.}
        \Indp
        \Repeat{\rm converged}{
            Sample $u$ from $\gU$ and let $\vx \gets \vx_u$.\\
            Randomly drop out $\vx$ to obtain $\vc$.\\
            Sample $t \sim \gU(1,T)$ and $\vz_t \sim \gN(\mF_t\vx, \sigma_t^2\bm{I})$.\\
            Take a gradient step on $\nabla_
                \theta\|\hat{\vx}_\theta(\vz_t, \vc, t) - \vx\|^2$.\\
        }
        \Indm
        \KwOut{Denoiser parameters $\theta$.}
        \Indp
    \end{algorithm}
    \DecMargin{1.5em}
}

\subsubsection{Denoiser Architecture}

\begin{figure}
    \centering
    \includegraphics[width=0.85\columnwidth]{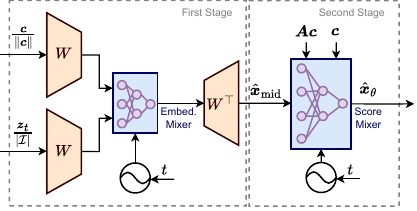}
    \caption{The architecture of the denoiser $\hat{\vx}_\theta$ in GiffCF.}\label{fig:denoiser}
    \Description{In the first stage, c divided by its norm, and z_t divided by the item count, are respectively fed into two identical encoders labeled W, and then, together with t processed by a sinusoidal encoder, are fed into an MLP labeled Embed. Mixer, which has 3 units in the input layer, 2 units in the hidden layer, and 1 unit in the output layer. The output is fed into a decoder labeled W^T to get x_mid. In the second stage, x_mid is fed into an MLP labeled Score Mixer, together with Ac, c, and t processed by another sinusoidal encoder, to output x_theta^hat. The MLP has 4 units in the input layer, 2 units in the hidden layer, and 1 unit in the output layer.}
\end{figure}

We adopt a two-stage design of the denoiser $\hat{\vx}_\theta$. (1) In the first stage, the network encodes both the latent preference signal $\vz_t$ and the historical interaction signal $\vc$ by multiplying them with a shared item embedding matrix $\mW \in \R^{|\gI|\times d}$ after proper normalization. The embeddings are then concatenated and fed into an item-wise MLP to mix information from the two sources. To account for timestep information, we also concatenate a timestep embedding to the input of the MLP. After that, the mixed embedding is decoded back into the interaction space by multiplying with the transposed item embedding matrix $\mW^\top$. This decoding operation can be seen as a dot product between the mixed user embedding and each item embedding in $\mW$, which is a common practice in matrix-factorization models \cite{korenMatrixFactorizationTechniques2009}. (2) In the second stage, we concatenate the intermediate score vector $\vx_{\text{mid}}$ from the first stage with the historical interaction vector $\vc$ and its smoothed version $\mA\vc$, and feed them into another item-wise MLP to predict the final interaction vector $\hat{\vx}_0$. Again, we concatenate a timestep embedding to the input of the MLP. The trainable parameters include the item embedding matrix $\mW$ and four MLPs: embedding mixer, score mixer, and two timestep encoders on the sinusoidal basis. The sizes of two mixers are shown in \Figref{fig:denoiser}. All the MLPs have one hidden layer with Swish activation. The overall architecture is illustrated in \Figref{fig:denoiser}. 

{
    \IncMargin{1.5em}
    \begin{algorithm}[t]
        \caption{GiffCF inference.}\label{alg:inference}
        \Indm
        \KwIn{$\theta$ and historical interactions $\vx_u$ of a user $u$.}
        \Indp
        Let $\vc \gets \vx_u$ and $\vz_T \gets \mF_T\vc$.\\
        \For{$t=T,\dots,1$}{
            Predict $\hat{\vx}_\theta \gets \hat{\vx}_\theta(\vz_t, \vc, t)$.\\
            Update $\vz_{t-1} \gets \mF_{t-1}\hat{\vx}_\theta + \sigma_{t-1}/\sigma_t(\vz_t - \mF_t\hat{\vx}_\theta)$.\\
        }
        Let $\vx \gets \vz_0$.\\
        \Indm
        \KwOut{Predicted preference scores $\vx$.}
        \Indp
    \end{algorithm}
    \DecMargin{1.5em}
}

\section{Experiments}

In this section, we conduct experiments on three real-world datasets to verify the effectiveness of GiffCF.
We aim to answer the following research questions:
\begin{itemize}[leftmargin=*]
    \item \textbf{RQ1} How does GiffCF perform compared to state-of-the-art baselines, including diffusion recommender models and graph signal processing techniques?
    \item \textbf{RQ2} How does the number of diffusion steps $T$, the smoothing schedule (controlled by $\alpha$), and the noise schedule (controlled by $\sigma_T$) affect the performance of GiffCF?
    \item \textbf{RQ3} How does the reverse process iteratively improve the recommendation results? Does our design of the denoiser architecture contribute to its performance?
\end{itemize}

\begin{table*}
    \setlength\extrarowheight{1pt}
    \centering\small
    \caption{\textbf{Performance comparison on all three datasets.} The best results are highlighted in bold and the second-best results are underlined. {\itshape\%Improv.} represents the relative improvements of GiffCF over the best baseline results. * implies the improvements over the best baseline are statistically significant ($p$-value $< 0.05$) under one-sample t-tests.}\label{tab:results}
    \begin{tabularx}{\textwidth}{l|l||*{8}{Y}|YY}
        \hline
        \textbf{Dataset}                       & \textbf{Metric}    & MF          & LightGCN           & Mult-VAE           & DiffRec            & L-DiffRec          & LinkProp           & GF-CF              & BSPM               & GiffCF              & {\itshape \%Improv.} \\
        \hline
        \multirow{6}{*}{\textbf{MovieLens-1M}} & \textbf{Recall@10} & 0.0885      & 0.1112             & 0.1170             & \underline{0.1178} & 0.1174             & 0.1039             & 0.1088             & 0.1107             & \textbf{0.1251}$^*$ & 6.17\%                      \\
                                               & \textbf{Recall@20} & 0.1389      & 0.1798             & 0.1833             & 0.1827             & \underline{0.1847} & 0.1509             & 0.1718             & 0.1740             & \textbf{0.1947}$^*$ & 5.44\%                      \\
                                               & \textbf{NDCG@10}   & 0.0680      & 0.0838             & 0.0898             & \underline{0.0901} & 0.0868             & 0.0852             & 0.0829             & 0.0838             & \textbf{0.0962}$^*$ & 6.76\%                      \\
                                               & \textbf{NDCG@20}   & 0.0871      & 0.1089             & \underline{0.1149} & 0.1148             & 0.1122             & 0.1031             & 0.1067             & 0.1079             & \textbf{0.1221}$^*$ & 6.28\%                      \\
                                               & \textbf{MRR@10}    & 0.1202      & 0.1363             & 0.1493             & \underline{0.1507} & 0.1394             & 0.1469             & 0.1383             & 0.1388             & \textbf{0.1577}$^*$ & 4.63\%                      \\
                                               & \textbf{MRR@20}    & 0.1325      & 0.1495             & 0.1616             & \underline{0.1630} & 0.1520             & 0.1574             & 0.1505             & 0.1513             & \textbf{0.1704}$^*$ & 4.53\%                      \\
        \hline
        \multirow{6}{*}{\textbf{Yelp}}         & \textbf{Recall@10} & 0.0509      & 0.0629             & 0.0595             & 0.0586             & 0.0589             & 0.0604             & 0.0630             & \underline{0.0630} & \textbf{0.0648}$^*$ & 2.92\%                      \\
                                               & \textbf{Recall@20} & 0.0852      & \underline{0.1041} & 0.0979             & 0.0961             & 0.0971             & 0.0980             & 0.1029             & 0.1033             & \textbf{0.1063}$^*$ & 2.12\%                      \\
                                               & \textbf{NDCG@10}   & 0.0301      & 0.0379             & 0.0360             & 0.0363             & 0.0353             & 0.0370             & 0.0381             & \underline{0.0382} & \textbf{0.0397}$^*$ & 3.88\%                      \\
                                               & \textbf{NDCG@20}   & 0.0406      & 0.0504             & 0.0477             & 0.0487             & 0.0469             & 0.0485             & 0.0502             & \underline{0.0505} & \textbf{0.0523}$^*$ & 3.59\%                      \\
                                               & \textbf{MRR@10}    & 0.0354      & 0.0446             & 0.0425             & 0.0445             & 0.0411             & 0.0445             & 0.0451             & \underline{0.0452} & \textbf{0.0476}$^*$ & 5.34\%                      \\
                                               & \textbf{MRR@20}    & 0.0398      & 0.0497             & 0.0474             & 0.0493             & 0.0460             & 0.0493             & 0.0501             & \underline{0.0503} & \textbf{0.0527}$^*$ & 4.84\%                      \\
        \hline
        \multirow{6}{*}{\textbf{Amazon-Book}}  & \textbf{Recall@10} & 0.0686      & 0.0699             & 0.0688             & 0.0700             & 0.0697             & \underline{0.1087} & 0.1049             & 0.1055             & \textbf{0.1121}$^*$ & 3.08\%                      \\
                                               & \textbf{Recall@20} & 0.1037      & 0.1083             & 0.1005             & 0.1011             & 0.1029             & \underline{0.1488} & 0.1435             & 0.1435             & \textbf{0.1528}$^*$ & 2.69\%                      \\
                                               & \textbf{NDCG@10}   & 0.0414      & 0.0421             & 0.0424             & 0.0451             & 0.0440             & \underline{0.0709} & 0.0688             & 0.0696             & \textbf{0.0733}$^*$ & 3.32\%                      \\
                                               & \textbf{NDCG@20}   & 0.0518      & 0.0536             & 0.0520             & 0.0547             & 0.0540             & \underline{0.0832} & 0.0807             & 0.0814             & \textbf{0.0858}$^*$ & 3.16\%                      \\
                                               & \textbf{MRR@10}    & 0.0430      & 0.0443             & 0.0455             & 0.0502             & 0.0468             & 0.0762             & 0.0750             & \underline{0.0763} & \textbf{0.0795}$^*$ & 4.15\%                      \\
                                               & \textbf{MRR@20}    & 0.0471      & 0.0487             & 0.0482             & 0.0540             & 0.0506             & 0.0807             & 0.0795             & \underline{0.0808} & \textbf{0.0842}$^*$ & 4.16\%                      \\
        \hline
    \end{tabularx}
\end{table*}

\subsection{Experimental Setup}

\subsubsection{Datasets}

For a fair comparison, we use the same preprocessed and split versions of three public datasets suggested by \cite{wangDiffusionRecommenderModel2023}:
\begin{itemize}[leftmargin=*]
    \item \textbf{MovieLens-1M} contains 571,531 interactions among 5,949 users and 2,810 items with a sparsity of 96.6\%.
    \item \textbf{Yelp} contains 1,402,736 interactions among 54,574 users and 34,395 items with a sparsity of 99.93\%.
    \item \textbf{Amazon-Book} contains 3,146,256 interactions among 108,822 users and 94,949 items with a sparsity of 99.97\%.
\end{itemize}

Each dataset is split into training, validation, and testing sets with a ratio of 7:1:2. For each trainable method, we use the validation set to select the best epoch and the testing set to tune the hyper-parameters and get the final results. To evaluate the top-$K$ recommendation performance, we report the average Recall@$K$ (normalized as in \cite{liangVariationalAutoencodersCollaborative2018}), NDCG@$K$ \cite{heTriRankReviewawareExplainable2015,heNeuralCollaborativeFiltering2017}, and MRR@$K$ \cite{chaeCFGANGenericCollaborative2018,wangVariableIntervalTime2021} among all the users, where we set $K\in\{10, 20\}$.

\subsubsection{Baselines}

The following baseline methods are considered:

\begin{itemize}[leftmargin=*]
    \item \textbf{MF} \cite{korenMatrixFactorizationTechniques2009} \& \textbf{LightGCN} \cite{heLightGCNSimplifyingPowering2020}, two representative recommender models that optimize the BPR loss \cite{rendleBPRBayesianPersonalized2009}. The latter linearly aggregates embeddings on the user-item bipartite graph.
    \item \textbf{Mult-VAE} \cite{liangVariationalAutoencodersCollaborative2018}, which generates interaction vectors from a multinomial distribution using variational inference.
    \item \textbf{DiffRec} \& \textbf{L-DiffRec} \cite{wangDiffusionRecommenderModel2023}, unconditional diffusion recommender models with standard Gaussian diffusion. The latter generates the embedding of an interaction vector encoded by grouped VAEs.
    \item \textbf{LinkProp} \cite{fuRevisitingNeighborhoodbasedLink2022} \& \textbf{GF-CF} \cite{shenHowPowerfulGraph2021} \& \textbf{BSPM} \cite{choiBlurringSharpeningProcessModels2023}, graph signal processing techniques for CF (see \Tabref{tab:filter}), which proved to outperform embedding-based models on large sparse datasets. The last one also simulates the heat equation with a sharpening process.
\end{itemize}

\subsubsection{Hyper-parameters}

For DiffRec and L-DiffRec, we reuse the checkpoints released by the authors, which are tuned in a vast search space. For the other baselines, we set all the user/item embedding sizes, hidden layer sizes, cut-off dimensions of ideal low-pass filters to 200. The dropout rates for Mult-VAE and all diffusion models are set to 0.5. Hyper-parameters not mentioned above are tuned or set to the default values suggested in the original papers. For our own model, by default, we set $\beta=\gamma=\delta=\frac{1}{2}$ for item-item similarities, $T=3$ as the number of diffusion steps, $\alpha=1.5$ as the smoothing strength, $\sigma_T = 0$ as the noise scale, and tune the ideal low-pass filters weight $\omega$ in $\{0.0, 0.1, 0.2, 0.3\}$, the noise decay $\frac{\sigma_{t-1}}{\sigma_t}$ in $\{0.1, 0.5, 1.0\}$. We optimize all the models using Adam \cite{kingmaAdamMethodStochastic2017} with a constant learning rate in $\{10^{-5},10^{-4},10^{-3}\}$. 

We implement GiffCF in Python 3.11 and Tensorflow 2.14. All the models are trained on a single RTX 3080 Ti GPU. For more details, please refer to our released code.

\subsection{Comparison with the State-of-the-Art}\label{sec:comparison}

The overall performance of GiffCF and the baselines on all three datasets is in \Tabref{tab:results}. The following observations can be made:

\begin{itemize}[leftmargin=*]
    \item All baseline models based on user/item embeddings (MF, LightGCN, Mult-VAE, DiffRec, L-DiffRec) show competitive performance on small datasets. However, on the larger and sparser Amazon-Book dataset, there is a significant performance drop compared to graph signal processing techniques (LinkProp, GF-CF, BSPM). This is because decomposing the high-dimensional interaction matrix into low-rank embedding matrices loses a lot of detailed information. As a result, these models struggle to capture the complex structure of the interaction space.
    \item When the number of parameters is comparable and training is sufficient, previous DMs for CF (DiffRec, L-DiffRec) do not exhibit a significant advantage over simpler baselines. This suggests that there is plenty of room for improvement in the design of DM and necessitates the introduction of more advanced techniques.
    \item GiffCF outperforms all the baselines on all the metrics on all the datasets. The superiority of GiffCF can be attributed to (1) the helpful prior knowledge of the interaction space structure introduced by advanced graph signal smoothing filters; (2) the hierarchical reconstruction of the interaction vector via refining and sharpening in the reverse process; (3) the two-stage denoiser effectively mixing different sources of information. We will analyze the effectiveness of each component in later sections.
\end{itemize}

\begin{figure}
    \includegraphics[width=\columnwidth]{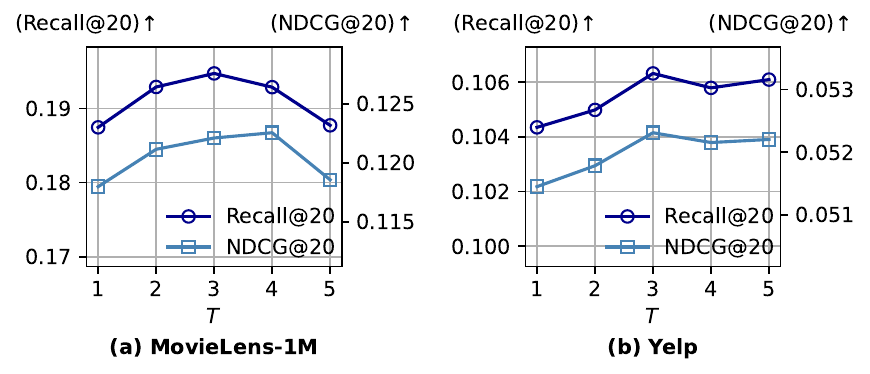}
    \caption{Effect of the number of diffusion steps $T$.}\label{fig:results_n_steps}
    \Description{Line graphs showing Recall@20 and NDCG@20 on (a) MovieLens-1M and (b) Yelp. The x-axis represents T from 1 to 5. Most metrics achieve the highest value at T=3, with a few oscillations or declines beyond 3 steps.}
\end{figure}

\begin{figure}
    \includegraphics[width=\columnwidth]{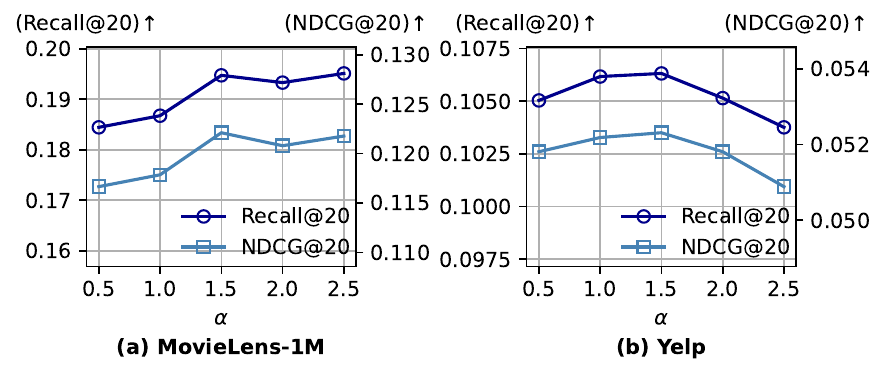}
    \caption{Effect of the smoothing strength $\alpha$.}\label{fig:results_alpha}
    \Description{Line graphs showing Recall@20 and NDCG@20 on (a) MovieLens-1M and (b) Yelp. The x-axis represents alpha from 0.5 to 2.5. Most metrics achieve the highest value at alpha=1.5, with a few oscillations or declines beyond 1.5.}
\end{figure}

\begin{figure}
    \includegraphics[width=\columnwidth]{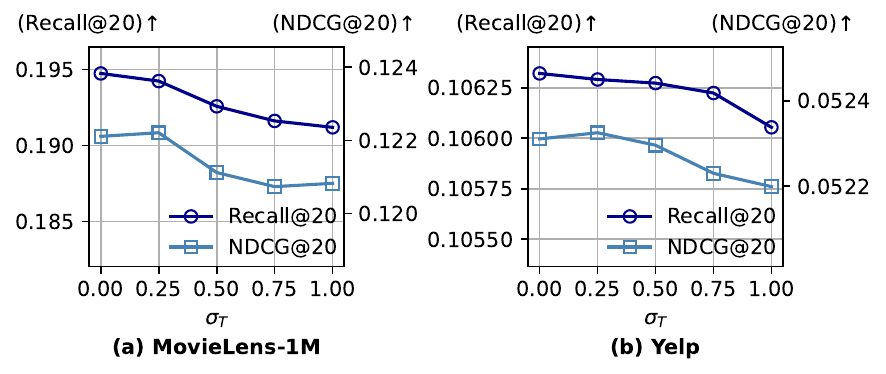}
    \caption{Effect of the noise scale $\sigma_T$.}\label{fig:results_noise_scale}
    \Description{Line graphs showing Recall@20 and NDCG@20 on (a) MovieLens-1M and (b) Yelp. The x-axis represents sigma_T from 0.0 to 1.0. Most metrics consistently decline as sigma_T increases, with only NDCG@20 at 0.25 slightly higher than at 0.0.}
\end{figure}

\subsection{Sensitivity on the Diffusion Schedule}

To demonstrate the effectiveness of graph signal diffusion and to provide guidance on scheduling-related hyper-parameter selection, we conduct sensitivity analysis on the number of diffusion steps $T$, the smoothing strength $\alpha$, and the noise scale $\sigma_T$.

\subsubsection{Effect of $T$}

We varied the number of diffusion steps $T$ from 1 to 5 and trained GiffCF from scratch. From the results in \Figref{fig:results_n_steps}, we can see that (1) as $T$ increases, the model performance first rises significantly, highlighting the advantage of a hierarchical generative model over those with a single latent variable. (2) beyond 3 steps, some performance metrics exhibit oscillations or declines, implying that an excessive number of sampling steps will not necessarily lead to higher accuracy. This bottleneck could stem from the limited expressiveness of the denoiser or potential inaccuracy in the prior estimation. Considering the computational burden associated with increasing steps, we suggest a uniform choice of $T=3$ to achieve near-optimal performance with acceptable execution time.

\subsubsection{Effect of $\alpha$}

The smoothing strength $\alpha$ describes the extent of information exchange in graph signal forward process, analogous to thermal diffusivity in a physical heat equation. To inspect its effect, we compare the results with $\alpha$ changing from 0.5 to 2.5 at an interval of 0.5. Quite similar to the effect of $T$, we observe that the performance first increases and then oscillates or declines as $\alpha$ exceeds 1.5, suggesting that a moderate smoothing strength is preferred. We interpret this phenomenon as follows: when $\alpha$ is below a certain threshold, $\vz_T$ may be under-smoothed and entangled with the original signal; when $\alpha$ is too large, $\vz_T$ may be over-smoothed and amplifying the difference between $\mA\vx$ and $\vx$ in an erroneous way. Both circumstances can impede the denoiser from discerning the hierarchical information within latent variables, ultimately leading to suboptimal performance. Based on experiments, $\alpha=1.5$ is a good choice overall.

\subsubsection{Effect of $\sigma_T$}

As discussed earlier, adding Gaussian noise in the diffusion process is not necessary for CF and poses a risk of information loss. To verify the effect of noise scale $\sigma_T$ on graph signal diffusion, we compare the results with $\sigma_T$ changing from 0.0 to 1.0 at an interval of 0.2. The results in \Figref{fig:results_noise_scale} clearly illustrate that with an increase in the noise scale, the model performance shows a declining trend. Hence, we suggest abandoning the noise term in the forward process. In this way, we allow the model to attend to and learn from clean and smoothed preference signals, making GiffCF a noise-free DM. It should be noted that even though we set the noise scale to zero, we still define the noise decay $\frac{\sigma_{t-1}}{\sigma_t}$ to control the refining term in graph signal reverse process. We will analyze the role of $\frac{\sigma_{t-1}}{\sigma_t}$ in the next section.

\subsection{Analysis of the Reverse Process}\label{sec:reverse}

\subsubsection{Role of the Refining Term}

\begin{figure}
    \includegraphics[width=\columnwidth]{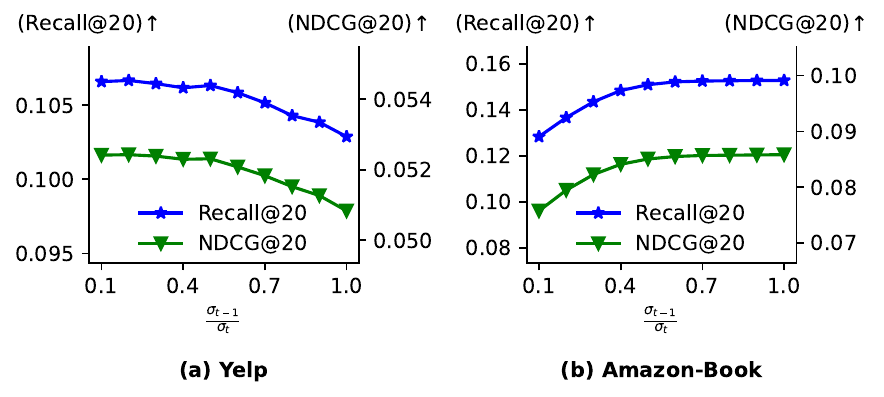}
    \caption{Effect of the noise decay $\frac{\sigma_{t-1}}{\sigma_t}$ (which controls the refining term in \Eqref{eq:refine-sharpen-decomposition}) during reverse sampling.}\label{fig:results_decay}
    \Description{Line graphs showing Recall@20 and NDCG@20 on (a) Yelp and (b) Amazon. The x-axis represents sigma_{t-1} divided by sigma_t from 0.1 to 1.0. On Yelp, the metrics asymptotically increase as the ratio decreases, while on Amazon, the metrics asymptotically increase as the ratio increases.}
\end{figure}

In \Secref{sec:refine-sharpen}, there is a so-called refining term that updates the previous latent signal $\vz_{t}$ toward a refined prediction $\mF_{t}\hat{\vx}_{\theta}$. To understand it from an empirical perspective, we change the ratio $\frac{\sigma_{t-1}}{\sigma_t}$ during inference and see how the model performs. The results are in \Figref{fig:results_decay}. On Yelp, we find that the smaller $\frac{\sigma_{t-1}}{\sigma_t}$, the better the performance. On Amazon-Book, the opposite trend is observed. In short, this is because embedding-based denoiser performs better on smaller datasets. In this case, setting $\frac{\sigma_{t-1}}{\sigma_t}$ to a small value can emphasize the refinement and mitigate the misspecification of the prior $\vz_T$. On larger and sparser datasets, preserving $\mF_T\vc$ as a prior estimate proves to be more beneficial, and hence setting $\frac{\sigma_{t-1}}{\sigma_t}$ near 1.0 is preferred.

\subsubsection{Quality of the Sampling Trajectory}

\begin{figure}
    \includegraphics[width=\columnwidth]{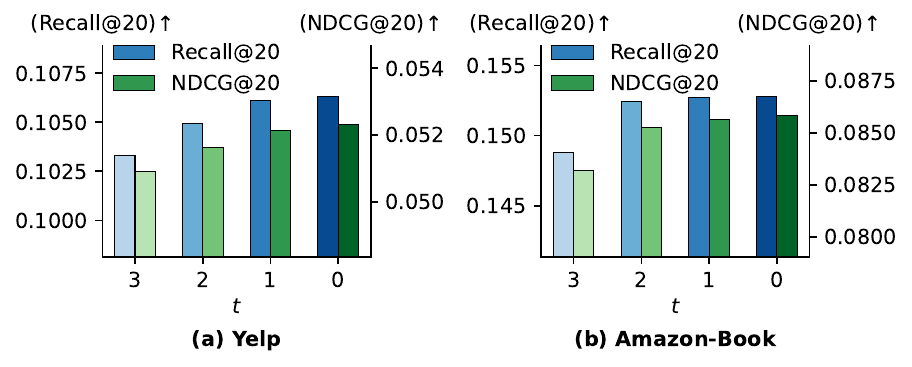}
    \caption{Recommendation performance using latent signals $\vz_t$ at each reverse sampling step $t$.}\label{fig:results_reverse}
    \Description{Bar charts showing Recall@20 and NDCG@20 on (a) Yelp and (b) Amazon. The x-axis represents the t from T=3 to 0. The metrics consistently increase as t decreases, with each step bringing less improvement than the previous one.}
\end{figure}

To show that the reverse process is indeed improving the preference scores through iterative refining and sharpening, we evaluate the quality of the sampling trajectory $\vz_t, t = T, \dots, 0$ in terms of the recommendation performance. The results in \Figref{fig:results_reverse} show that $\vz_t$ progressively produces better results after each sampling step, which validates the effectiveness of graph signal reverse process.

\subsubsection{Ablation on the Denoiser Architecture}

Another important finding is that the denoiser architecture plays a crucial role in the performance of GiffCF. To verify this, we conduct ablation studies by either removing $\vz_t$ in the first stage (w/o latent), removing $\vc$ in the first stage (w/o pre-cond), or removing $\mA\vc$ and $\vc$ in the second stage (w/o post-cond). The results in \Tabref{tab:ablation} show that both components are indispensable for achieving the best performance, consistent with our intuition. On the one hand, the information from both $\vz_t$ and $\vc$ can encourage learning a better embedding matrix $\mW$ that utilizes the prior knowledge from the forward process. On the other hand, feeding $\mA\vc$ and $\vc$ into a final item-wise mixer can help the denoiser remove the artifacts caused by low-rank compression. This is especially important for high-dimensional datasets, where a solely embedding-based model is prone to underfitting. To sum up, the two-stage denoiser architecture effectively mixes different sources of information from $\vz_t$ and $\vc$.

\begin{table}
    \centering\small
    \caption{Ablation studies on the denoiser architecture. The best results are highlighted in bold.}\label{tab:ablation}
    \begin{tabularx}{\columnwidth}{l||l|YYY}
        \hline
        \textbf{Dataset}                       & Arch. of  $\hat{\vx}_\theta$          & Recall@20          & NDCG@20          & MRR@20           \\
        \hline
        \multirow{3}{*}{\textbf{MovieLens-1M}} & -                                     & \textbf{0.1947}    & \textbf{0.1221}  & \textbf{0.1704}  \\
                                               & w/o latent                            & 0.1667             & 0.1105           & 0.1639           \\
                                               & w/o pre-cond                          & 0.1829             & 0.1159           & 0.1648           \\
                                               & w/o post-cond                         & 0.1888             & 0.1204           & 0.1698           \\
        \hline
        \multirow{3}{*}{\textbf{Yelp}}         & -                                     & \textbf{0.1063}    & \textbf{0.0523}  & \textbf{0.0527}  \\
                                               & w/o latent                            & 0.1055             & 0.0520           & 0.0522           \\
                                               & w/o pre-cond                          & 0.1041             & 0.0509           & 0.0512           \\
                                               & w/o post-cond                         & 0.0851             & 0.0418           & 0.0431           \\
        \hline
        \multirow{3}{*}{\textbf{Amazon-Book}}  & -                                     & \textbf{0.1528}    & \textbf{0.0858}  & \textbf{0.0842}  \\
                                               & w/o latent                            & 0.1506             & 0.0846           & 0.0827           \\
                                               & w/o pre-cond                          & 0.1518             & 0.0850           & 0.0829           \\
                                               & w/o post-cond                         & 0.1057             & 0.0570           & 0.0556           \\
        \hline
    \end{tabularx}
\end{table}

\section{Related Works}

The related works can be classified into two categories: DM for recommendation and DM with general corruption.

\paragraph{DM for recommendation.}

As a powerful family of generative models, DM has recently garnered attention in the field of recommender systems. Most of these models rely on Gaussian noise to explore either the interaction or the user/item embedding space, encompassing the subfields of CF \cite{walkerRecommendationCollaborativeDiffusion2022,wangDiffusionRecommenderModel2023}, sequential recommendation \cite{wangConditionalDenoisingDiffusion2023,liDiffuRecDiffusionModel2023,duSequentialRecommendationDiffusion2023,yangGenerateWhatYou2023,liuDiffusionAugmentationSequential2023}, \emph{etc}. An exception to this trend is the use of discrete diffusion for multi-stage recommendation \cite{linDiscreteConditionalDiffusion2023}. For CF, non-Gaussian diffusion in the interaction space is still under-explored. Notably, \citet{choiBlurringSharpeningProcessModels2023} designed a graph signal processing technique inspired by DM, which also involves a sharpening process to improve recommendation performance. Our work stands out as the first to incorporate graph signal processing into the theoretical framework of DM for CF, thereby achieving the best of both worlds.

\paragraph{DM with general corruption.}

Researchers have been exploring general forms of corruption in DM. For example, \citet{bansalColdDiffusionInverting2023} explored cold DM without noise; \citet{darasSoftDiffusionScore2022} explored DM with general linear operators; \citet{rissanenGenerativeModellingInverse2022} and \citet{hoogeboomBlurringDiffusionModels2022} explored DM with image blurring filters; \citet{austinStructuredDenoisingDiffusion2021} explored discrete DM with general transition probabilities; \emph{etc}. We are the first to propose DM with graph smoothing filters for implicit feedback data, contributing to a general corruption form for graph-structured continuous state spaces.

\section{Conclusion and Future Work}

In this work, we present GiffCF, a novel DM for CF that smooths and sharpens graph signals by simulating the heat equation. The proposed hierarchical generative model effectively leverages the adjacency of the item-item similarity graph and enables high-quality reconstruction of user-item interactions. Extensive experiments not only demonstrate the superiority of our method compared to the state-of-the-art, but also provide empirical insights into the design of several key components. We believe that GiffCF can serve as a strong baseline for future recommender systems research.

To extend our work, one can consider: (1) designing more powerful denoisers or training strategies to accelerate convergence and enhance performance; (2) integrating other conditional information, such as user demographics, to better control the reverse process; (3) adapting our model to other types of recommendation tasks, probably by exploring similar corruption forms in the embedding space. Hopefully, our work could inspire more researchers to think beyond the boundaries of conventional image DMs and design task-specific DMs with general corruption. By that means, we can truly unleash the potential of DM in information retrieval.

\begin{acks}
    This work was partially supported by National Natural Science Foundation of China (Grant No.92370204), Guangzhou-HKUST(GZ) Joint Funding Program (Grant No.2023A03J0008), Education Bureau of Guangzhou Municipality, and Guangdong Science and Technology Department. It was also partially supported by China Postdoctoral Science Foundation (Grant No.2023M730785) and China Postdoctoral Science Foundation (Grant No.2023M741849).
\end{acks}

\bibliographystyle{ACM-Reference-Format}
\bibliography{refs}

\end{document}